\documentclass{article}
\usepackage{spconf,amsmath,graphicx}
\usepackage{epstopdf}
\usepackage{slashed}
\usepackage{bm}
\usepackage{mdwmath}
\usepackage{amsthm}
\usepackage{url}
\usepackage{amssymb}
\usepackage{enumerate}
\usepackage{algorithm} 
\usepackage{algorithmic} 
\usepackage{subfigure}
\usepackage{appendix}
\usepackage{cite}
\usepackage{makecell,rotating,multirow,diagbox}
\usepackage{color}

\graphicspath{{figures/}}

\def\f{{\mathbf f}}

\def\v{{\mathbf v}}

\def\x{{\mathbf x}}
\def\y{{\mathbf y}}

\def\0{{\mathbf 0}}

\def\I{{\mathbf I}}

\def\D{{\mathbf D}} 
 
\def\L{{\mathbf L}}

\def\V{{\mathbf V}}
\def\W{{\mathbf W}}
\def\X{{\mathbf X}}

\def\cG{{\mathcal G}}

\def\bLambda{{\boldsymbol \Lambda}}

\title{Graph Neural Net using Analytical Graph Filters and Topology Optimization for Image Denoising}
\name{{Weng-tai Su$^{\dag}$, Gene Cheung$^{\ddag}$, Richard Wildes$^{\ddag}$ and Chia-Wen Lin$^{\dag}$}
\thanks{The work of G. Cheung and R. Wildes was supported in part by the Natural Sciences and Engineering Research Council of Canada (NSERC).}
\thanks{The work of W.-t. Su was supported in part by the Ministry of Science and Technology, Taiwan, under Grants 108-2917-I-007-013.}} 

\address{$^{\dag}$Department of Electrical Engineering, National Tsing Hua University, Taiwan\\
$^{\ddag}$Department of Electrical Engineering \& Computer Science, York University, Canada
}

\begin{document}
\ninept
\pagestyle{plain}
\maketitle
\begin{abstract}
While convolutional neural nets (CNNs) have achieved remarkable performance for a wide range of inverse imaging applications, the filter coefficients are computed in a purely data-driven manner and are not explainable. 
Inspired by an analytically derived CNN by Hadji et al., in this paper we construct a new layered graph neural net (GNN) using GraphBio as our graph filter. 
Unlike convolutional filters in previous GNNs, our employed GraphBio is analytically defined and requires no training, and we optimize the end-to-end system only via learning of appropriate graph topology at each layer. 
In signal filtering terms, it means that our linear graph filter at each layer is always intrepretable as low-pass with known biorthogonal conditions, while the graph spectrum itself is optimized via data training.
As an example application, we show that our analytical GNN achieves image denoising performance comparable to a state-of-the-art CNN-based scheme when the training and testing data share the same statistics, and when they differ, our analytical GNN outperforms it by more than 1dB in PSNR.
\end{abstract}

\section{Introduction}
The advent of deep learning based methods such as \textit{convolutional neural nets} (CNNs) has brought about a seismic paradigm shift in how inverse imaging systems, such as image denoising \cite{vemulapalli2016deep,zhang2017beyond}, super resolution \cite{dong2014learning, kim2016accurate} and deblurring \cite{tao2018scale}, are designed and built. 
Though these inverse imaging problems are traditionally solved by signal processing tools like filters \cite{milanfar13}, wavelets \cite{rioul91} and sparse dictionaries \cite{tosic11} derived analytically based on mathematical models of images \cite{chang2000adaptive,dabov2007image, gu2014weighted,pang2017graph
}, they are now unceremoniously discarded and replaced by data-driven neural nets trained using large collections of labelled data. 
While there is no denying the supreme performance of these trained CNNs, open fundamental questions about their operations remain: 
i) Are all the degrees of freedom afforded by thousands of network parameters necessary to achieve good performance? 
ii) How to best train a CNN if only a small collection of labelled data is available?
iii) If the statistics of the training and testing data differ significantly---a statistical mismatch---to what extent would the performance of the trained CNN be affected?

In this paper, we investigate these issues using a novel \textit{graph neural net} (GNN)\footnote{A GNN differs from a CNN in that the convolutional filter in each layer is a graph filter operating on an irregular data kernel described by a graph.} architecture where the employed convolutional filters are entirely analytically defined.
Our work is inspired by \cite{hadji2017spatiotemporal}, where fixed Gaussian filters---requiring zero data training---are combined with point-wise non-linearity and pooling operators to compose each convolutional layer, resulting in an ``explainable" CNN that nonetheless achieves state-of-the-art performance in image texture recognition. 
Analogously, we choose an analytical graph filter---a biorthogonal graph wavelet called \textit{GraphBio} \cite{narang2013compact} in our implementation---to build each graph convolutional layer, which we stack together to build a GNN.
Unlike \cite{hadji2017spatiotemporal}, we perform data training to optimize edge weights in an 8-connected graph to filter each pixel patch, so that the graph spectrum can be data-adaptive.

Compared to recent works in graph spectral image processing \cite{cheung2018graph} that extend from the rapid development of the \textit{graph signal processing} (GSP) field \cite{ortega18ieee}, a key difference in our work is that the construction of the underlying graph for pixel patch processing is not ad-hoc (\textit{e.g.}, using bilateral filter weights \cite{tomasi98}) but data-trained. 
One exception is \cite{zeng2019deep}, where edge weights of a graph are learned before a denoising problem using a \textit{graph Laplacian regularizer} (GLR) prior is solved.
While the solution to the quadratic programming problem in \cite{zeng2019deep} can also be interpreted as a low-pass graph filter, the solution requires solving a system of linear equations, which is complex even if a fast method like \textit{conjugate gradient} (CG) \cite{powell1977restart} is used. 
In contrast, our work simply implements an analytical graph wavelet as the convolutional filter, which is known to be fast in execution.

Our experiments show the following. 
First, compared to a state-of-the-art CNN-based image denoising algorithm DnCNN \cite{zhang2017beyond}, our GNN has comparable performance when sufficient data are available for training. 
This shows that analytical graph filters combined with just enough degrees of freedom for graph learning are sufficient to achieve good denoising performance.
Second, when the statistics between training and testing data differ, our GNN can outperform DnCNN by more than 1dB in PSNR.
This demonstrates that with fewer degrees of freedom only for data-driven graph learning, our GNN is less likely to overfit compared to DnCNN.

The outline of the paper is as follows. 
We first review important fundamentals in GSP and graph wavelets in Section\;\ref{sec:prelim}. 
We then motivate and describe our designed graph neural net in Section\;\ref{sec:arch}.
We argue that our proposed GNN benefits from guaranteed filter stability in Section\;\ref{sec:stable}.
Finally, we present our experimental results and conclusion in Section\;\ref{sec:results} and \ref{sec:conclude}, respectively.

\section{Preliminaries}
\label{sec:prelim}
\subsection{GSP Definitions}

We first define basic definitions in GSP to facilitate understanding of our proposed GNN.
A graph $\cG$ with $M$ nodes can be specified using an \textit{adjacency matrix} $\W \in \mathbb{R}^{M \times M}$, where $w_{ij} > 0$ connects nodes $i$ and $j$. 
$w_{ij}=0$ implies there is no edge between nodes $i$ and $j$.
The \textit{degree matrix} $\D$ is a diagonal matrix with diagonal terms $d_{ii} = \sum_{j=1}^{M} w_{ij}$. 
The \textit{graph Laplacian matrix} $\L$ is simply computed as $\L=\D-\W$.
One can eigen-decompose $\L = \V \bLambda \V^{\top}$, where $\bLambda = \mathrm{diag}(\lambda_1, \ldots, \lambda_M)$ are the eigenvalues and $\V = [\v_1, \ldots, \v_M]$ are the eigenvectors. 
One can show via Gershgorin Circle Theorem (GCT) that if the undirected graph contains only non-negative edge weights, then graph Laplacian $\L$ is positive semi-definite (PSD)  \cite{cheung2018graph}.
Given that PSD $\L$ has eigenvalues $\lambda_i \geq 0, \forall i$, eigen-pairs $(\lambda_i, \v_i)$ define the \textit{graph frequencies} (\textit{graph spectrum}) of graph $\cG$.

\subsection{Overview of GraphBio}

We overview a previously designed graph wavelet called GraphBio, which we employ as the analytical graph filter in our GNN architecture.
GraphBio is a critically sampled biorthogonal graph wavelet.
It is grounded in the fact that the \textit{spectral folding} phenomenon (well understood in regular data kernel when downsampling by $2$) is also observed on \textit{bipartite graphs} when samples of one of the partites are removed.
Operating on a bipartite graph, GraphBio then employs a partite removal operator that replaces the conventional ``downsample by 2" operator, and designs low-pass / high-pass filters to enable perfect reconstruction during synthesis.
Because of this design, when deploying GraphBio on general graphs that are not bipartite, a \textit{bipartite graph approximation} step is typically inserted before GraphBio prefiltering. Bipartite graph approximation for general graphs has been studied alone as a research topic \cite{zeng17}. 
Because we operate on a 8-connected pixel graph, finding an appropriate bipartite graph is significantly easier.
See Section \ref{sec:graph} for details.

\section{Architecture Design}
\label{sec:arch}
\begin{figure*}[!hbt]
	\centering
	\includegraphics[width=0.7\textwidth]{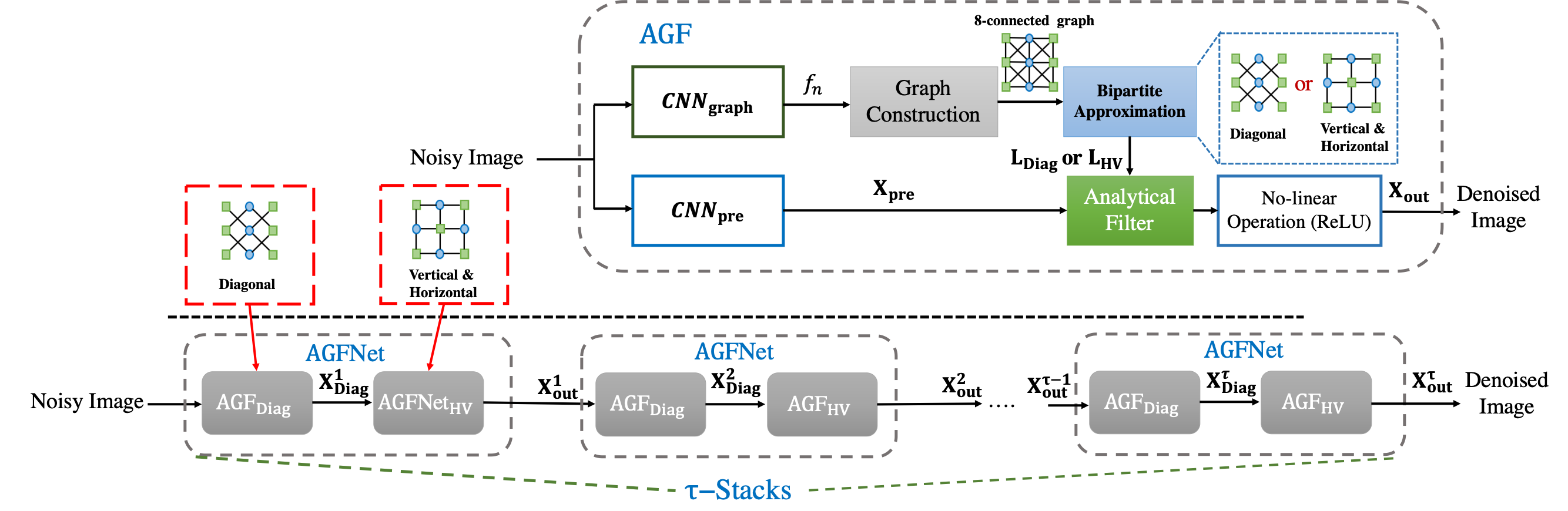}
	\vspace{-0.08in}
	\caption{Block diagram of the overall DeepAGF framework. Top:Block diagram of the proposed AGF, which uses analytical graph filters (diagonal or vertical / horizontal) for image denoising. Buttom: Block diagram of the $\tau$ stacks DeepAGF framework.}
	\label{fig:archfig}	
\end{figure*}

\begin{figure*}[!ht]
    \vspace{-0.01in}
	\centering
	\includegraphics[width=0.5\textwidth]{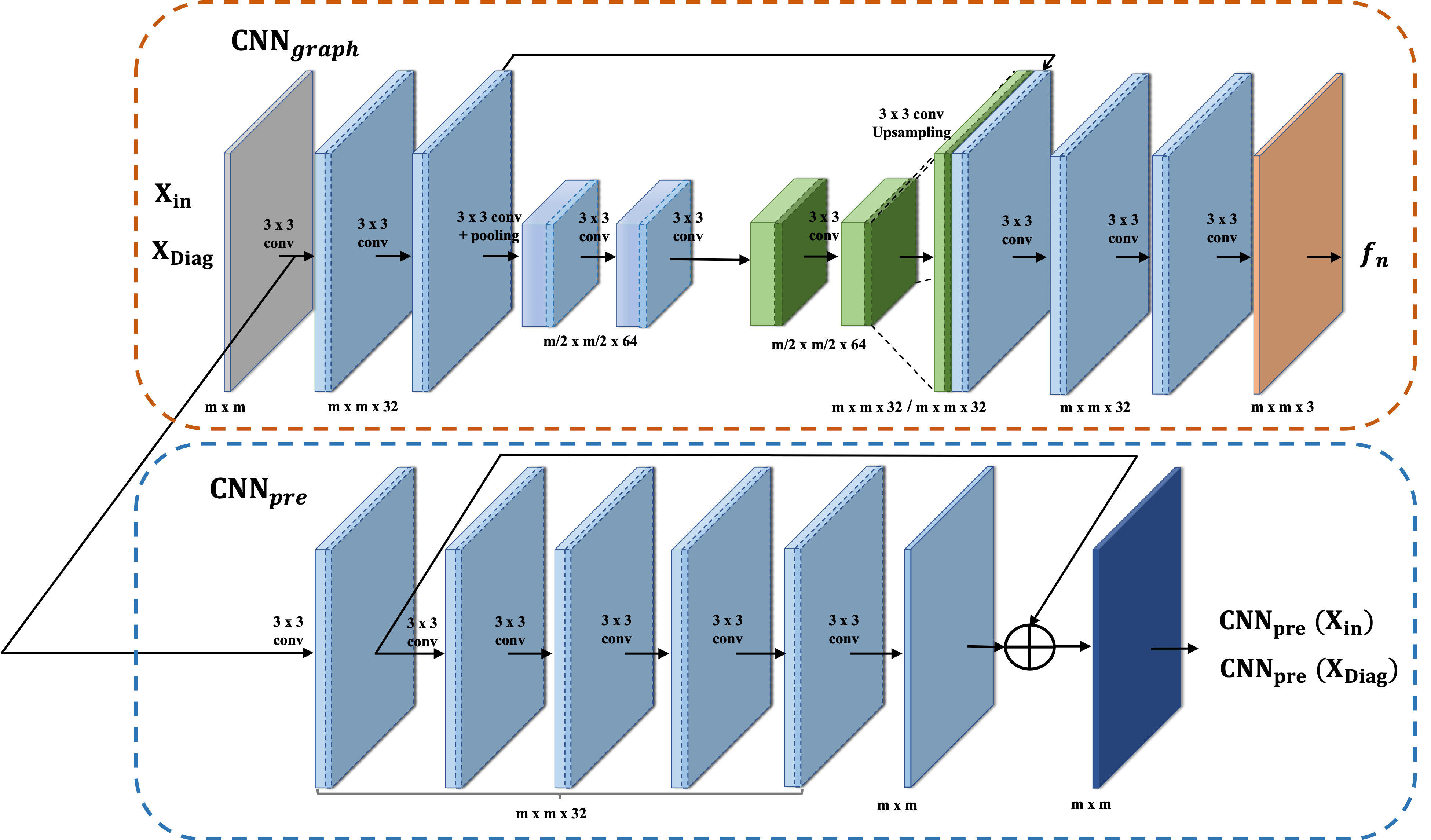}
	\vspace{-0.1in}
	\caption{Network architectures of $ \mathbf{CNN}_{graph}$ and $\mathbf{CNN}_{pre}$.}
	\label{fig:flowchart}	
	\vspace{-0.1 in}
\end{figure*}

We design a GNN with a chosen analytical graph filter as the key building block.  
Fig.\,\ref{fig:archfig} depicts the block diagram (AGFNet) of our architecture.
It contains two CNNs: i) $\mathbf{CNN_{graph}}$ constructs an underlying graph, and ii) a lightweight CNN \cite{he2016deep} pre-filters the noisy image, similar to \cite{chatterjee2011patch}, prior to graph filtering using our chosen analytical filter. 
A notable feature in our architecture is that while the chosen analytical graph filter is fixed, $\mathbf{CNN_{graph}}$ learns the underlying graph. 
Given a learned graph,
we partition its edges into two bipartite graphs for separate graph filtering, where the pre-filtered output from $\mathbf{CNN_{pre}}$ is the input for the analytical graph filter. 
Finally, we employ a non-linear operation (ReLu \cite{krizhevsky2012imagenet}) to obtain the output from analytical graph filter.

\subsection{Graph Construction}


To reduce computation complexity, we first divide the input noisy image $\mathbf{X_{in}}$ into $K$ non-overlapping $m \times m$ pixel patches (\textit{i.e.}, $\X_{\mathbf{in}}^k \in \mathbb{R}^M$, $1\leq k \leq K$, $M=m \times m$) for individual processing, as done in \cite{hu2015graph, liu2014progressive,pang2017graph}.
The output of $\mathbf{CNN_{graph}}$ are feature matrices $\{{\f^k}\}^K_{k=1}$ (\textit{i.e.} $\f^k \in \mathbb{R}^{M \times N}$, $1\leq k \leq K$) corresponding to the $K$ patches, which are $N$-dimensional feature vectors $\f^k_i \in \mathbb{R}^N$ for each pixel $i$ in patch $k$. 
For each patch, we construct a graph $\cG$ to connect pixels in the patch for graph filtering. 
$\cG$ is chosen to be an 8-connected graph, and each edge weight $w^k_{ij}$ for  $k$-th patch is computed as follows:
\vspace{-0.1in}
\begin{align}
w^k_{ij} = \exp\left(
-\frac{\mathrm{dist}(k, i, j)}{2\epsilon^2}
\right),  1\leq k \leq K,
\label{eq:wig1}
\end{align}
where $\mathrm{dist}(k, i,j)$ is the \textit{feature distance} between nodes $i$ and $j$ for $k$-th patch.   
We compute feature distance $\mathrm{dist}(k, i,j)$ using the two corresponding feature vectors $\f^k_i$ and $\f^k_j$ as follows
\begin{align}
\mathrm{dist}(k, i,j) = \left( \f^k_i - \f^k_j \right)^{\top}
\left( \f^k_i - \f^k_j \right),  1\leq k \leq K,
\label{eq:wig2}
\end{align}
Note that using \eqref{eq:wig1} to compute $w^k_{ij}$ means that the edge weights are always non-negative, and thus $\L$ is guaranteed to be PSD---and the graph spectrum is well defined---as discussed in Section\,\ref{sec:prelim}.


\subsection{Graph Laplacian and Bipartite Approximation}
\label{sec:graph}

 


As discussed, GraphBio can operate only on bipartite graphs that exhibit the spectral folding phenomenon when one partite is discarded \cite{narang2013compact}.
However, the learned graph $\cG$ using $\mathbf{CNN_{graph}}$ is not bipartite.
Hence we must first partition the edges in $\cG$ into two or more bipartite graphs before using GraphBio for graph filtering. 
Because the learned graph is $8$-connected, we can easily separate the edges into two bipartite graphs: i) vertical / horizontal, and ii) two diagonal directions, as shown in Fig.\,\ref{fig:archfig}. 
We employ GraphBio on bipartite graphs specified by Laplacian matrices $\{\L_{\mathbf{Diag}}^k\}^K_{k=1}$ and $\{\L_{\mathbf{HV}}^k\}^K_{k=1}$ corresponding to diagonal and horizontal / vertical edge bipartite graphs respectively. 

\subsection{Repeated Analytical Graph Filter}

Using analytical graph filter (AGF), we compute its output, image denoised patches $\X_{\mathbf{out}}^k$, as
	\begin{align}
	\X_{\mathbf{out}}^k &= \sigma[F(\L_{\mathbf{HV}}^k,
	\mathbf{CNN}_{\mathbf{pre}} (\X_{\mathbf{Diag}}^k))],  1\leq k \leq K,
	\label{eq:AFmodel1} \\
	\X_{\mathbf{Diag}}^k &= \sigma[F(\L_{\mathbf{Diag}}^k,
	\mathbf{CNN}_{\mathbf{pre}} (\X_{\mathbf{in}}^k))],  1\leq k \leq K,
	\label{eq:AFmodel2}
	\end{align}
where $\{\X_{\mathbf{Diag}}^k\}^K_{k=1}$ are the output of the GraphBio filter $F(.)$ using $\{\L_{\mathbf{Diag}}^k\}^K_{k-1}$ as Laplacian matrices and  $\{\mathbf{CNN}_{\mathbf{pre}} (\X_{\mathbf{in}}^k)\}^K_{k=1}$ as input,
and  $\{\X_{\mathbf{out}}^k\}^K_{k=1}$ are the output of the GraphBio filter $F(.)$ using $\{\L_{\mathbf{HV}}^k\}^K_{k-1}$ as Laplacian matrices and  $\{\mathbf{CNN}_{\mathbf{pre}} (\X_{\mathbf{Diag}}^k)\}^K_{k=1}$ as input.
$\{\mathbf{CNN}_{\mathbf{pre}} (\X_{\mathbf{Diag}}^k)\}^K_{k=1}$ are the pre-filtered results using $\mathbf{CNN_{pre}}$, and $\sigma(.)$ is the non-linear operation ReLU after the graph filter. 
Finally, the denoised image $\mathbf{\X_{out}}$ is obtained by concatenating the denoised patches $\{\X_{\mathbf{out}}^k\}^K_{k=1}$. 
The diagram of the analytical graph filter (AGFNet) is shown in Fig.\,\ref{fig:archfig}. 
Each block includes two sub-blocks of AGF (\textit{i.e.}, AGF$_{\mathbf{Diag}}$ and AGF$_{\mathbf{HV}})$, and each sub-block includes two CNN models, graph construction, bipartite approximation, analytical filter and non-linearity operation.

To achieve effective denoising, classic literature \cite{elad2006image, milanfar13, dabov2007image} filters the noisy image iteratively to gradually enhance the image quality. 
Similar to previous work \cite{vemulapalli2016deep}, we employ repeated filtering by cascading $\tau$ blocks of AGFNet. 
To effectively learn the AGFNet modules in the cascading structure, we share the same CNN parameters ($\mathbf{CNN_{graph}}$ and $\mathbf{CNN_{pre}}$) for all cascaded blocks. 
Our proposed AGFNet iteratively performs denoising $\tau$ times to obtain a final denoised image, $\mathbf{X_{out}^{\tau}}$, as shown in Fig.\,\ref{fig:archfig}.
Based on this repeated filter architecture, the objective function of DeepAGFNet framework can easily be defined as the \textit{mean squared  error} (MSE) between $\mathbf{\X_{gt}}$ and $\mathbf{\X_{out}^{\tau}}$:
	\begin{align}
	\mathbf L_{MSE}(\mathbf{\X_{gt}},\mathbf{\X_{out}^{\tau}}) = &\frac{1}{HW}\sum_{i=1}^{H}\sum_{j=1}^{W}(\mathbf{\X_{gt}}(i,j)-\mathbf{\X_{out}^{\tau}}(i,j))^2,
	\label{eq:loss}
	\end{align}
where $\mathbf{\X_{gt}}$ is ground-truth image, $\mathbf{\X_{out}^{\tau}}$ is the denoised image, $H$ is the height and $W$ is the width of the image. 

\subsection{Network Architecture}

The two different CNN models used in our architecture are shown in details in Fig.\,\ref{fig:flowchart}. 
For $\mathbf{CNN_{graph}}$, we adopt a fully-convolutional encoder-decoder architecture with skip connections \cite{ronneberger2015u}, including two deconvolutional layers, and the output channels of $\mathbf{CNN_{graph}}$ are set to 3 ($N=3$) to construct the graphs.
Similar to \cite{zeng2019deep}, we employ a residual structure \cite{he2016deep} as generated by 6 convolution layers to build the pre-filter $\mathbf{CNN_{pre}}$.

\section{Filter Stability}
\label{sec:stable}
We argue that one advantage of using an analytical graph filter in our GNN instead of a data-trained one is the guaranteed stability of the filter.
As a comparison point,  \cite{zeng2019deep} solves the denoising problem 
\begin{align}
\min_{\x} \left\| \y - \x \right\|_2^2 + \mu \x^{\top} \L \x
\end{align}
where $\y$ is the noisy observation, $\x^{\top} \L \x$ is GLR, and $\mu$ is a parameter that trades off the fidelity term and GLR.
The solution $\x^*$ is computed by solving:
\vspace{-0.0in}
\begin{align}
    \left( \I + \mu \L \right) \x^* = \y
    \label{eq:linSys}
\end{align}
The stability of the linear system \eqref{eq:linSys} depends on the \textit{condition number}---ratio of the largest to smallest eigenvalues $\lambda_{\max} / \lambda_{\min}$ of the coefficient matrix $\I + \mu \L$. 
It was shown to be bounded via GCT \cite{varga2010gervsgorin} for a 8-connected graph with maximum edge weight $1$.

Analogously, in our GraphBio implementation, the filter response given spectrum $\{\lambda_1, \ldots, \lambda_m\}$ is also guaranteed to be stable \cite{narang2013compact}.
Specifically, given each node has maximum degree of $8$ (maximum $8$ connected edges each with maximum weight $1$), one can show that $\lambda_{\max}$ of graph Laplacian $\L$ is also upper-bounded via GCT.
Since GraphBio is approximated by a polynomial function of $\L$ via Chebyshev approximation, stability of GraphBio depends on the matrix norm of $\L$, which is $\lambda_{\max}$. 
Since $\lambda_{\max}$ is bounded as discussed, one can conclude also that our graph filter is also stable no matter what graph is learned from our CNN implementation.

\section{Experimentation}
\label{sec:results}
\begin{figure*}[!hbt]
\centering
\includegraphics[width=1.0\textwidth]{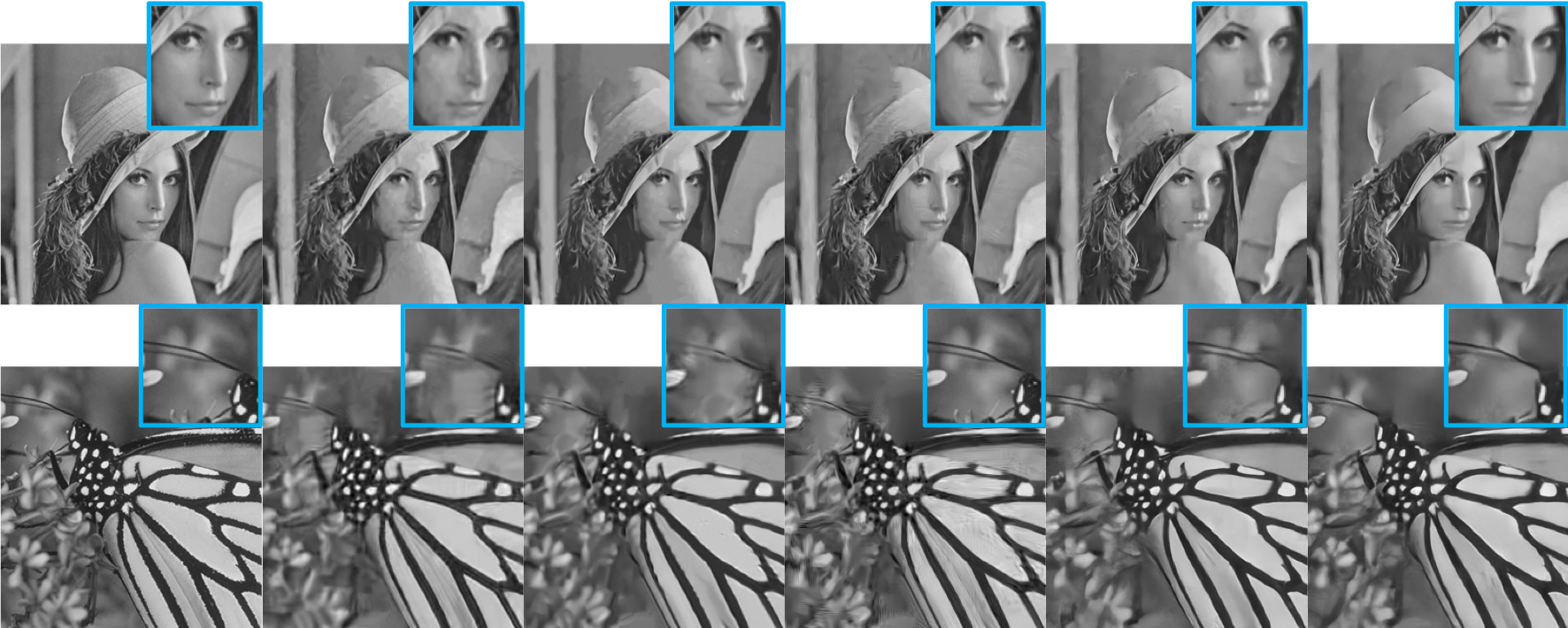}
\vspace{-0.25in}
\caption{Denoising results for Lena and Monarch, from left to right: noisy level $\sigma=50$, Original, BM3D, WNNM, OGLR, DnCNN-S, DeepAGF.} 
\label{fig:comp}	
\end{figure*}

We compare our proposed GNN against several state-of-the-art denoising schemes.
The competing schemes are two model-based methods (BM3D \cite{dabov2007image} and WNNM \cite{gu2014weighted}), a graph-based method (OGLR \cite{pang2017graph}) and a state-of-the-art deep learning model for image denoising (DnCNN \cite{zhang2017beyond}).

\subsection{Experimental Setup}

We first test the addition of independent and identically distributed (i.i.d.) additive white Gaussian noise (AWGN), where we train our proposed DeepAGF for denoising with a high noisy variance $\sigma=50$. 
We use the dataset (with 400 gray-scale images of size 180 $\times$ 180) provided by \cite{zhang2017beyond} for training. During the training phase, the noisy images, accompanied by their ground-truth images, are fed to the network for training. 
The denoising performance is evaluated on 12 commonly used test images (\textit{i.e.}, Set12) with sizes of 256 $\times$ 256 or 512 $\times$ 512, similarly done in \cite{zhang2017beyond}. 
For objective evaluation, peak signal-to-noise ratio (PSNR) is employed. 
For our proposed DeepAGF, we set the patch size to 24 $\times$ 24 (\textit{i.e.}, $m=24^2=576$), train the network on 74k patches and set the batch size to 32 for 200 epochs. 
We use two cascades of AGFNet for our proposed DeepAGF. 

	\begin{table}
		\vspace{-0.05in}
		\begin{center}
			\begin{scriptsize}
			\scriptsize
				\begin{tabular}{|c|c|c|c|c|c|} \hline
				    \multirow{2}{*}{Name} & \multicolumn{5}{c|}{Method      $\space(\sigma=50)$} \\
				    \cline{2-6}
					         &BM3D \cite{dabov2007image}   &WNNM \cite{gu2014weighted}   &OGLR \cite{pang2017graph}  &DnCNN-S \cite{zhang2017beyond}   &DeepAGF\\ \hline\hline
					Cman     &26.15  &26.42  &25.93  &26.99     &26.74\\ \hline
					House    &29.66	 &30.44  &29.4   &30.15     &30.02\\ \hline
					Peppers  &26.69	 &26.93  &26.55  &27.24     &27.16\\ \hline
					Starfish &24.93	 &25.36  &24.8   &25.73     &25.54\\ \hline
					Monarch  &25.78	 &26.17  &25.62  &26.86     &26.90\\ \hline
					Airplane &25.03	 &25.36  &24.97	 &25.92     &25.67\\ \hline
					Parrot   &25.81	 &26.09  &25.78  &26.49     &26.33\\ \hline
					Lena     &29.05	 &29.23  &28.84	 &29.34     &29.37\\ \hline
					Barbara  &27.21	 &27.78  &27.13	 &26.20     &25.89\\ \hline
					Boat     &26.64	 &26.88  &26.58	 &27.13     &27.06\\ \hline
				    Man      &26.81	 &26.85  &26.62  &27.18     &27.15\\ \hline
					Couple   &26.47	 &26.64  &26.38  &26.81     &26.78\\ \hline\hline
           \textbf{Avg.}      &26.69	 &27.01  &26.55  &27.17     &27.05\\ \hline
				\end{tabular}
			\end{scriptsize}
		\end{center}
		\vspace{-0.2in}
		\caption{ Set12 PSNR (dB)}
		\label{tab:t1}
	\end{table}
	    \begin{table}%

	\begin{center}
	    \vspace{-0.1in}
		\begin{scriptsize}
			\begin{tabular}{|c|c|c|}\hline 
			   &DnCNN-S  &DeepAGF\\ \hline
                Parameters  &0.55 M & 0.32M\\ \hline

			\end{tabular}
		\end{scriptsize}
	\end{center}
	\vspace{-0.2in}
	\caption{Parameter count comparison for different methods}
	\label{tab:t2}
	\vspace{-0.2in}
	\end{table}
	
    \begin{table}
	\begin{center}
	    \vspace{-0.2in}
		\begin{scriptsize}
			\begin{tabular}{|c|c|c|} \hline
			   &DnCNN-S  &DeepAGF\\ \hline\hline
                Avg. PSNR &18.46 &19.78\\ \hline

			\end{tabular}
		\end{scriptsize}
	\end{center}
	\vspace{-0.2in}
	\caption{Denoising results for train/test mismatch case ($\sigma=50$ and $\sigma=70$ for train and test) for Set12}
	\label{tab:t3}
	\vspace{-0.2in}
	\end{table}

\subsection{Quantitative Comparisons}

Table\,\ref{tab:t1} shows the average PSNR values of different denoising methods for 12 test images. 
Although our DeepAGF method is not the best, the CNN architecture employs only six layers for pre-filtering, which is small compared to the top performing DnCNN \cite{zhang2017beyond} that employs 17 layers. 
Further, our proposed DeepAGF achieves better performance than two model-based methods (BM3D and WNNM) and graph-based method (OGLR). 
We note that PSNR does not fully reflect image quality.
To demonstrate visual quality,  we also show a visual comparison of the denoising methods in Fig.\,\ref{fig:comp}. 
We observe that the DeepAGF provides the best visual quality: there are fewer artifacts and smoother results without loosing important detail (\textit{i.e.}, facial area for lena, Monarch's tentacles and part of the edge of the image).
Table\,\ref{tab:t2} shows our scheme's parameter count compared to DnCNN---our GNN saves more than $40 \%$ parameters.

    \begin{figure}
	\centering
	\vspace{-0.15in}
	\includegraphics[width=0.35\textwidth]{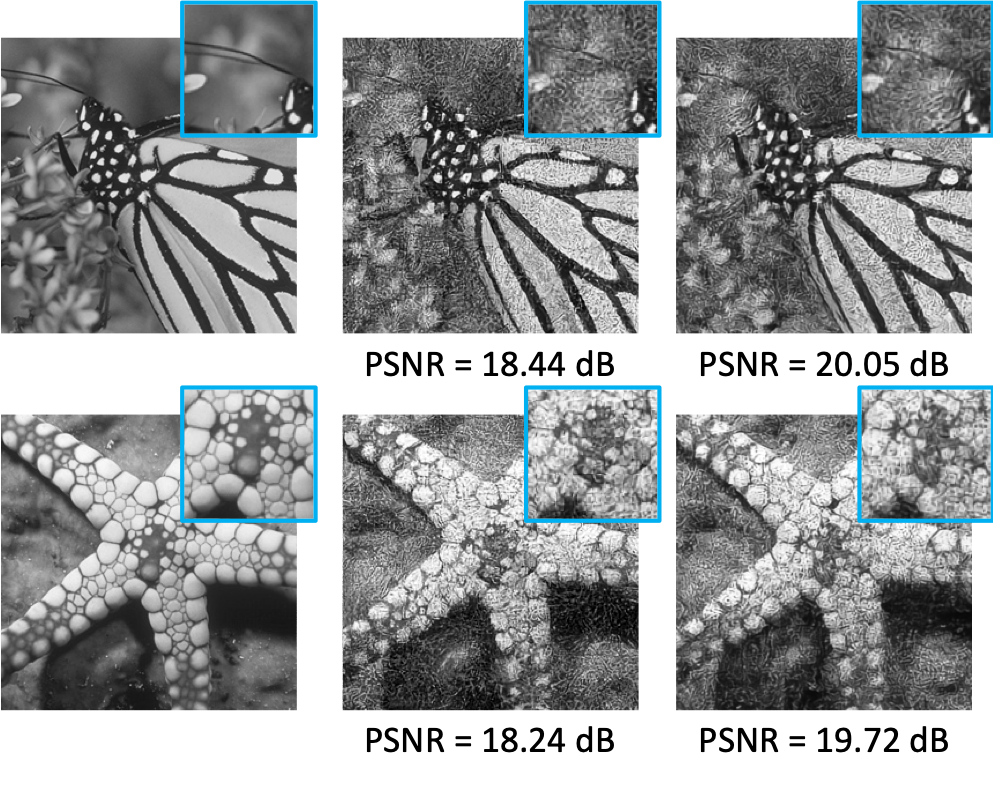}
	\vspace{-0.2in}
	\caption{Denoising results for Monarch and Starfish, from left to right: noisy level $\sigma=70$, Original, DnCNN-S, DeepAGF.} 
	\label{fig:comp2}	
    \end{figure}

For the case of statistical mismatch between training and testing data, we set $\sigma=50$ for training and set $\sigma=70$ for testing. 
Table\,\ref{tab:t3} shows the average PSNR values for our scheme and DnCNN-S. 
We observe that our analytical GNN outperforms DnCNN-S by more than 1dB in PSNR, demonstrating that our GNN is more robust to statistical mismatch.  
We also include a visual comparison for the mismatch case in Fig.\,\ref{fig:comp2}.

\section{Conclusion}
\label{sec:conclude}
We propose a new graph neural net (GNN) architecture for image denoising that employs an analytical graph wavelet filter---biorthogonal GraphBio in our implementation---while the underlying graph is optimized in a data-driven manner.
Compared to conventional CNNs, our architecture offers fewer degrees of freedom only for graph learning, while enjoying state-of-the-art denoising performance. 
Fewer degrees of freedom translates to a smaller likelihood to overfit.
We demonstrate this by showing that, when the statistics between training and testing data differ, our GNN outperforms competing CNNs by more than 1dB in PSNR.

\bibliographystyle{IEEEtran}
\bibliography{ref}
\end{document}